\documentclass[preprint,12pt]{elsarticle}
\usepackage{graphicx}
\usepackage{amsmath}   
\usepackage{etoolbox}
\usepackage{amssymb}
\usepackage{epsfig}
\usepackage{float}
\usepackage{lineno, hyperref}
\usepackage{url}
\usepackage{amsthm}
\usepackage[export]{adjustbox}
\usepackage[font=scriptsize,labelfont=bf,skip=5pt]{caption} 
\usepackage[belowskip=-25pt,aboveskip=2pt]{caption}
\usepackage{subfig}
\usepackage{xpatch}
\usepackage{hyperref}
\hypersetup{
    colorlinks=true,
    linkcolor=blue,
    filecolor=magenta,      
    urlcolor=cyan,
    pdftitle={ATLAS RPC Hermetic Sealing},
    pdfpagemode=FullScreen,
    }

\def\pmbanner{}
\begin{document}
\begin{frontmatter}
\title{ \pmbanner {Quality assurance tests and techniques to investigate and improve hermeticity of ATLAS Phase II Resistive Plate Chamber detectors}}
\author{Aashaq Shah\corref{cor1}}
\cortext[cor1]{Corresponding author \\ 
Email address: \href{mailto:aashaq.shah@cern.ch}{aashaq.shah@cern.ch} (Aashaq Shah) \\ 
On behalf of the ATLAS Muon Community}

\address{Cavendish Laboratory, University of Cambridge, Cambridge, United Kingdom}

\begin{abstract}

The Phase II upgrade of the ATLAS Muon Spectrometer will involve the installation of approximately 1000 next-generation single Resistive Plate Chamber (RPC) detectors, also referred to as singlets. This upgrade aims to enhance detector coverage, increase hit efficiency, and improve timing precision, ultimately strengthening the precision and robustness of the muon trigger system.  The upgrade is essential for maintaining the performance of the muon spectrometer in the high-luminosity environment of the HL-LHC, where increased radiation and event rates pose significant challenges. Currently, detector production is underway, with gas gaps being produced in Italy. To ensure the integrity of these gas gaps, especially their mechanical properties and gas tightness, several investigative techniques have been proposed and implemented.

This study discusses the results of Thermal Cycling tests performed on ATLAS RPC gas gaps. These tests were conducted between $-33^{\circ}$C and  $+35^{\circ}$C in a climate room at University of Cambridge and $-20^{\circ}$C and $+30^{\circ}$C in laboratories at INFN Frascati, providing insights into the effects of mechanical stress due to thermal expansion on gas leaks. Additionally, it outlines various methods employed to assess and enhance the gas tightness of the detectors. The ultimate objective is to produce hermetic RPC detectors, minimising the emission of environmentally harmful gases and mitigating their impact on global warming.
\end{abstract}

\end{frontmatter}

\section{Introduction}
\begin{figure}[H]
\centering
\includegraphics[width=8.5cm]{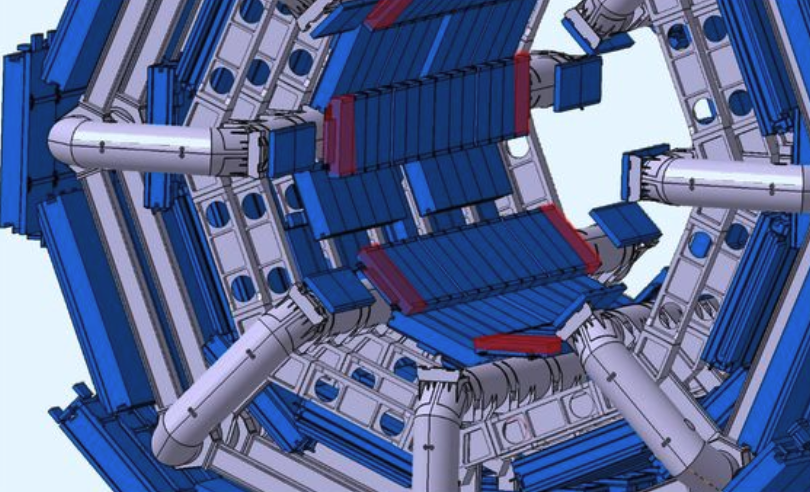}
\vspace{5pt}
\caption{The ATLAS Muon Spectrometer. 
The blue regions indicate the locations of the Resistive Plate Chambers (RPCs) and Monitored Drift Tube (MDT) chambers, while the grey areas depict the toroidal solenoid magnet and the overall mechanical structure of the experiment~\cite{Ref_BISFPGA}.}
\label{fig:atlas_muon_spectrometer}
\vspace{15pt}
\end{figure}

The ATLAS Muon Spectrometer, shown in Figure~\ref{fig:atlas_muon_spectrometer}, is undergoing upgrades as part of the Phase II upgrade program, in preparation for the high-luminosity phase of the Large Hadron Collider (HL-LHC)~\cite{Ref_ATLAExp, Ref_PhaseII_ATLASTDR}. A key component of this upgrade is the enhancement of the Resistive Plate Chamber (RPC) system, which involves installing over 300 Phase II RPC chambers. This includes 226 Barrel Inner (BI) chambers and 80 Barrel Outer (BO) and Middle (BM) chambers, which are designed to handle rates of up to 10 kHz/cm$^{2}$ in the high-radiation environment of the HL-LHC~\cite{Ref_PhaseIIMuonTDR}.

\begin{figure}[H]
\centering
\includegraphics[width=7.5cm]{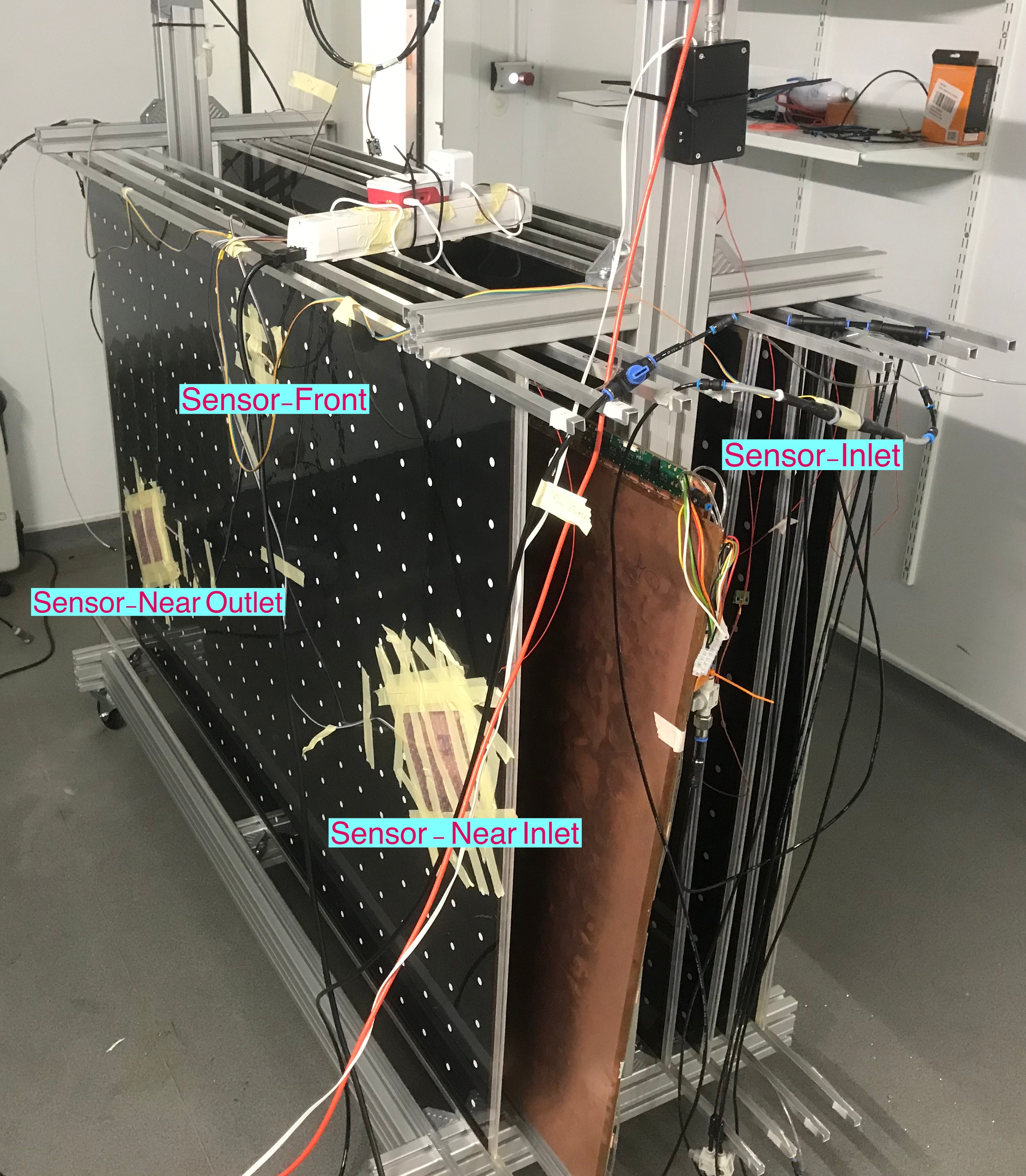}
\vspace{5pt}
\caption{Gas gaps, five in total, plus one complete Resistive Plate Chamber detector, placed in a climate chamber room at University of Cambridge, with multiple sensors to monitor their conditions. The Inlet and Outlet sensors are positioned inside the gas gap tubing at entry and exit points to measure internal gas conditions directly. Near Inlet and Near Outlet sensors (with the Outlet sensor not visible in this view) are externally attached to gas gap close to the inlet and outlet and are covered to monitor its conditions at these locations. Additionally, Front and Back sensors (Back sensor not visible) are externally attached to the gas gap to assess the climate room ambient conditions and track environmental variables across its surface.}
\label{fig:Camb_thermal_setup}
\vspace{15pt}
\end{figure}

The new chambers are expected to significantly enhance the performance of ATLAS Muon system. By increasing the RPC-based muon trigger system from six to nine layers, the robustness of the system will improve, and the overall detector acceptance will rise from 78\% to 92\%, reaching up to 96\% when combining Barrel Inner (BI) and Barrel Outer (BO) coincidences. The introduction of these new RPCs, featuring a thin gas gap (1 mm), will enable a timing resolution of approximately 0.4 ns for BI-RPCs and an improvement in legacy RPCs from 2 ns to 1.1 ns with upgraded readout electronics, thereby enhancing the accuracy of time-of-flight (ToF) measurements.

\begin{figure}[H]
\centering
\includegraphics[width=9cm]{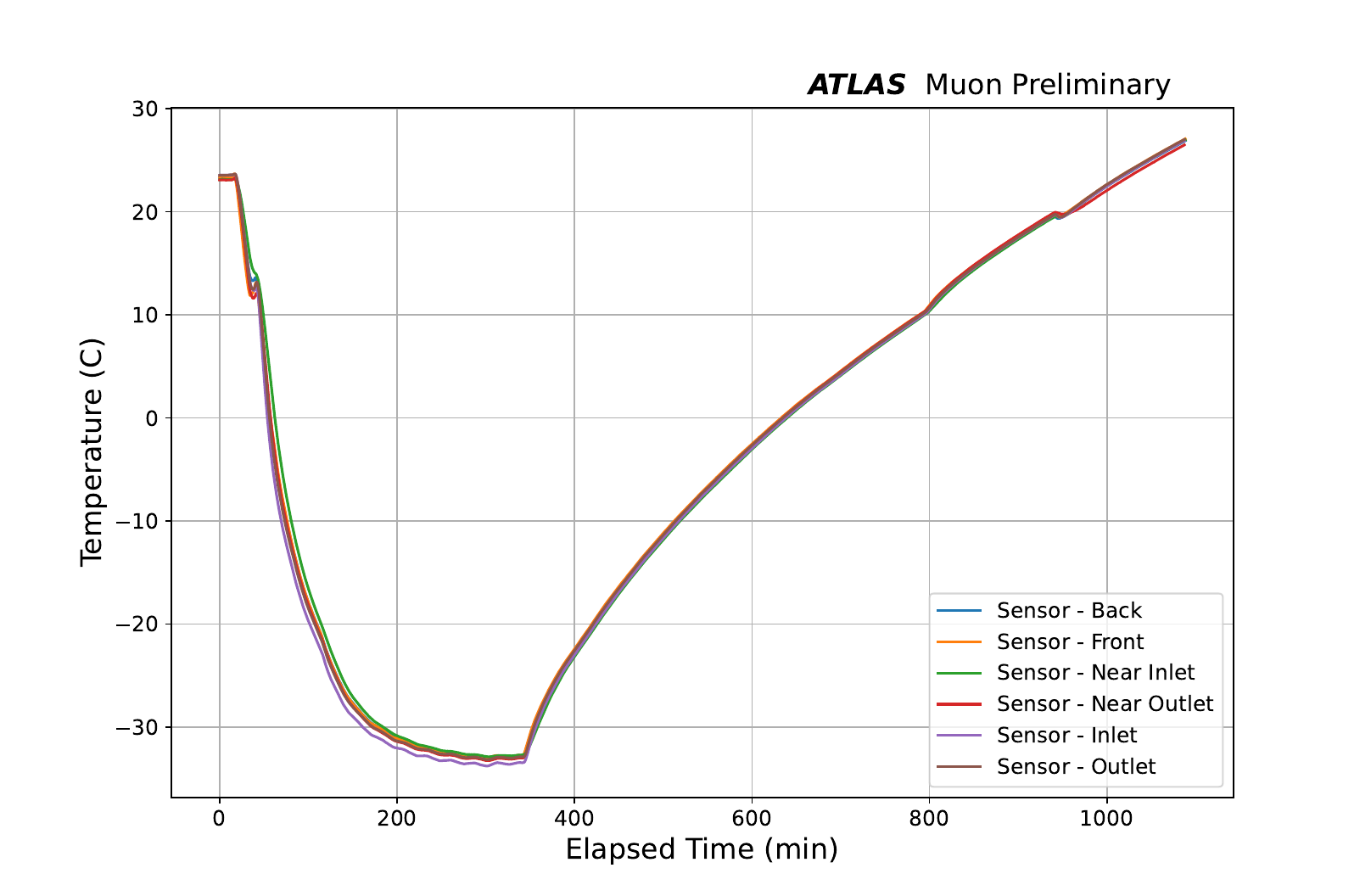}
\includegraphics[width=9cm]{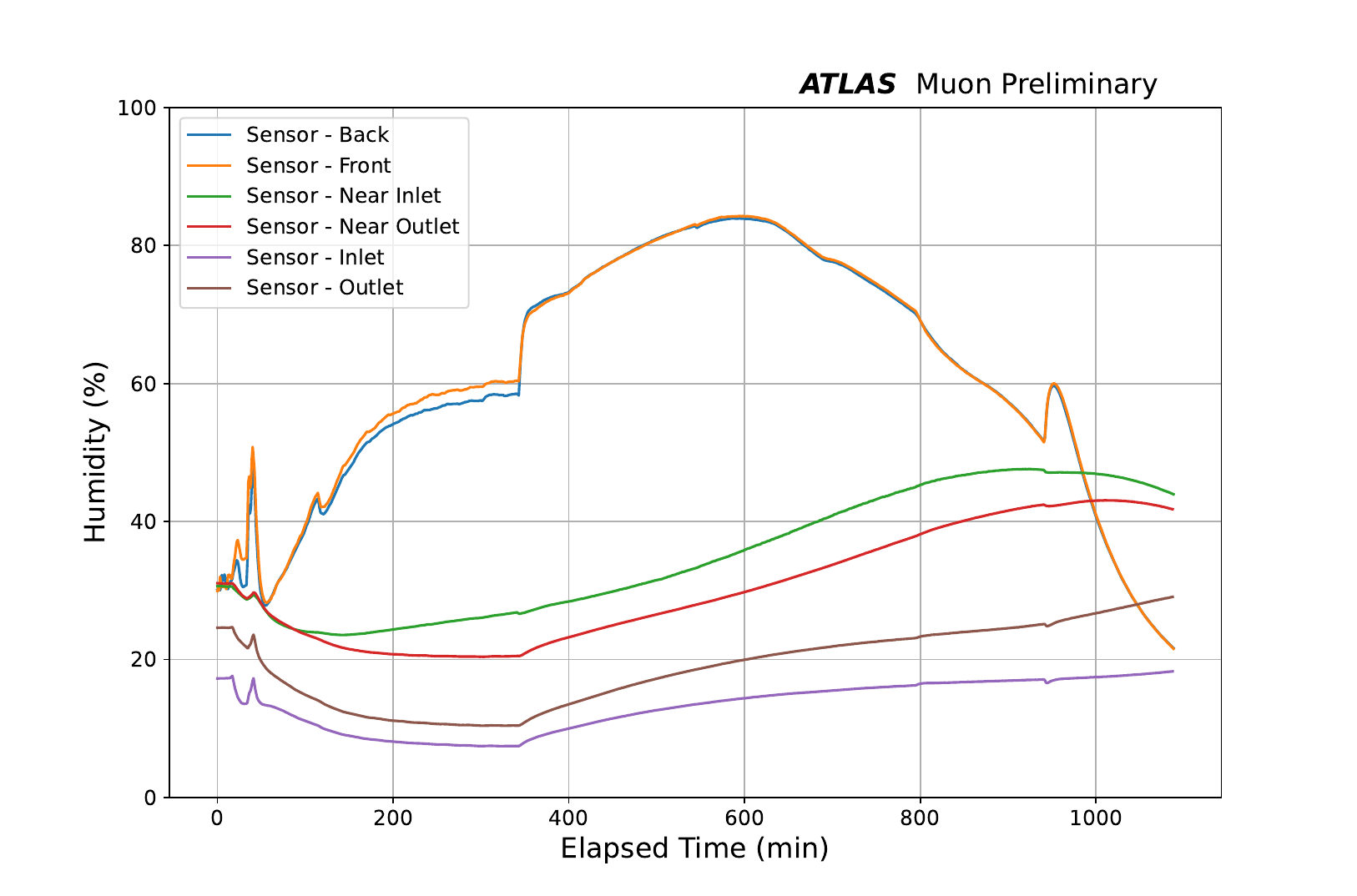}
\vspace{5pt}
\caption{Example of a thermal cycle performed in the Cambridge climate chamber over several hours, with recorded temperature conditions~(top). Sensors at multiple positions -- including within the gas gap inlet and outlet -- track the temperature response in relation to the separately controlled climate chamber conditions. Additional sensors on the front and back surfaces provide a broader view of the climate room ambient conditions. (bottom) Recorded relative humidity conditions during the same thermal cycle within the gas gap inlet and outlet, as well as positions near the inlet and outlet on the outside of the gas gap.}
\label{fig:Camb_thermal_cycling}
\vspace{15pt}
\end{figure}

The gas gaps used in these RPCs are primarily produced by General Tecnica Engineering (GTE) in Italy (with other potential production sites being considered in Germany and China)~\cite{Ref_BI_RPCPrdouction}. Each gas gap consists of a gas gap sandwiched between two High-Pressure Laminate (HPL) bakelite plates with a thin external graphite coating. After production, these gas gaps undergo several quality assurance tests, including surface resistivity measurements~\cite{Ref_RPC_PhasII_Production}. 
The study presented in this paper specifically focuses on evaluating the integrity and durability of these gas gaps under mechanical stress -- a key aspect of assessing reliability over a timespan of more than a decade during the operation of the HL-LHC.

In this study, we describe our Quality Assurance (QA) tests for the gas gaps of Phase II RPC. 
The thermal cycling (TC)\footnote{A thermal cycle refers to one complete variation between a positive temperature and a negative temperature, followed by a return to the initial temperature.} technique is central to these QA tests.
This technique applies mechanical stress to the gas gaps by exploiting variations in the thermal expansion coefficients of the constituent materials of the gas gaps through systematic temperature variations. 
The key advantage of this technique is that it applies stresses in a controlled manner and that these stresses are focused on the interfaces between the different constituents of the gas gap like e.g., high-pressure laminate (bakelite) and glue, where mechanical failure is most likely to occur.
Another advantage of this technique is that it can simulate the ageing of the gap in an environment with relatively stable conditions over a decade-long timespan on a short timescale of a couple of days by using a sufficiently large window for the temperature excursions in the TC protocol.
Similar TC procedures are used by ATLAS in the quality control and QA of the  Inner TracKer (ITk) strip modules~\cite{Ref_ATLAS_ITk_thermalQC} and serve as an inspiration for our tests.
However, unlike for the ITk that looks for mechanical failure modes like module cracking, our tests aim to reveal the mechanical resilience and potential failure points of the gas gaps in view of gas tightness, providing valuable insights into their structural integrity over time. 
The work presented here has supported improvements in gas gap design, thereby reducing emissions of RPC gases through leaks that could potentially develop over time during RPC operation and contribute to global warming.

Our QA tests through TC were carried within the climate-controlled chambers at both University of Cambridge and the INFN Frascati laboratories. 
These tests were conducted under conditions of down to $-33^{\circ}$C and up to $+35^{\circ}$ at Cambridge and down to $-20^{\circ}$C and up to $+20^{\circ}$C at Frascati.
Between the TCs, the gas gaps were  monitored for their mechanical resilience and gas tightness, and tests were conducted to identify potential leak points.


\section{Methodology and Test Setup} \label{sect:MechanicalStress}
Thermal cycling tests were conducted to assess the gas gaps under varying temperature conditions. As part of the procedure, gas gaps were initially evaluated for existing leaks by inflating them to an overpressure of either 3 or 5 mbar. The gas gaps are rated for up to 10 mbar overpressure, they were typically tested at 3 mbar. The use of 5 mbar in addition to 3 mbar was intended to expose small leaks that may not be detectable within a few minutes at a lower overpressure of 3~mbar, as will be clarified in subsequent discussions. After inflation, the pressure drop within the gas gaps was monitored using Kane 3200 Differential Pressure Manometer over a three-minute period. 
Based on empirical data from a sample of gas gaps at CERN, acceptable values for pressure drop were set at $\Delta$P $<$ 0.01 mbar per minute, establishing a benchmark for gas tightness. Following this initial leak assessment, thermal cycling tests were performed, and additional leak rate measurements were taken between thermal cycles to evaluate any changes in the integrity of the gas gaps.

In addition to manually measuring leak rates with a pressure manometer, a semi-automatic setup was developed to track pressure drop over time within the gas gaps. This setup was built using a Raspberry Pi and Bosch BME280 sensors~\cite{Ref_BOSCH_Sensor}, which can simultaneously measure pressure, temperature, and humidity. Two sensors were placed within the gas pipes at the inlet and outlet to monitor internal conditions within the gas gap, and two additional sensors were attached externally to the gas gap, positioned near the gas inlet and outlet. To avoid capturing ambient conditions, these external sensors were insulated with three layers of bubble wrap, ensuring they accurately monitored the conditions of the gas gap itself rather than ambient air during thermal cycling.

Furthermore, two additional sensors were placed on the front and back faces of one of the gas gaps, uncovered, to measure ambient conditions within the climate room and across the gas gap. For clarity, the sensors are labelled as Inlet, Outlet, Near Inlet, Near Outlet, Front, and Back, as shown in the setup in Figure~\ref{fig:Camb_thermal_setup}. This terminology is consistently used throughout the results presented in this study.

Figure~\ref{fig:Camb_thermal_setup} shows the experimental setup within the climate chamber room at the University of Cambridge. This setup includes five ATLAS Phase II Barrel Inner Small (BIS)-type gas gaps and one complete RPC detector, all placed in a climate-controlled environment designed to facilitate TC. The chamber room, equipped with two compressors, allows temperature variations from $-40^{\circ}$C to $+35^{\circ}$C, though for our tests, it was operated within a range of $-33^{\circ}$C to $+35^{\circ}$C. 
An example of a thermal cycle between $-33^{\circ}$C and $+25^{\circ}$C, conducted using this setup, is shown in Figure~\ref{fig:Camb_thermal_cycling} (top). Here, it can be observed that all sensors closely track the climate room's temperature variations. Meanwhile, Figure~\ref{fig:Camb_thermal_cycling} (bottom) illustrates the relative humidity at various points on one of the gas gaps. The sensors labelled Inlet and Outlet are housed in 3D-printed fittings connected to gas lines, recording internal gas conditions within the gap. Notably, the inlet sensor shows lower humidity than the outlet, as dry CO$_{2}$ gas was continuously flushed through the gas gap, and the outlet sensor captures a slight humidity increase due to exposure to the chamber’s humid environment. This small exchange of moisture with ambient air is further verified by the Near Inlet and Near Outlet sensors. These sensors start with similar humidity levels; however, the Near Outlet sensor shows a gradual decrease in humidity as it interacts with the gas flow, which becomes more humid as it picks up moisture within the gas gap. The Front and Back sensors provide additional readings of the external humidity surrounding the gas gap and within the broader climate room environment. 

\begin{figure}[H]
\centering
\includegraphics[width=8.5cm]{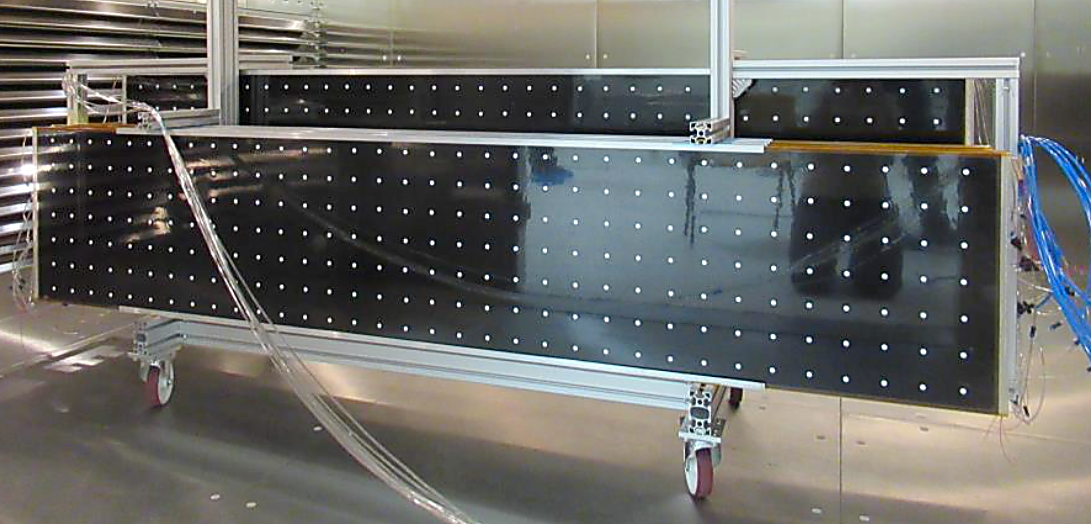}
\includegraphics[width=8.5cm]{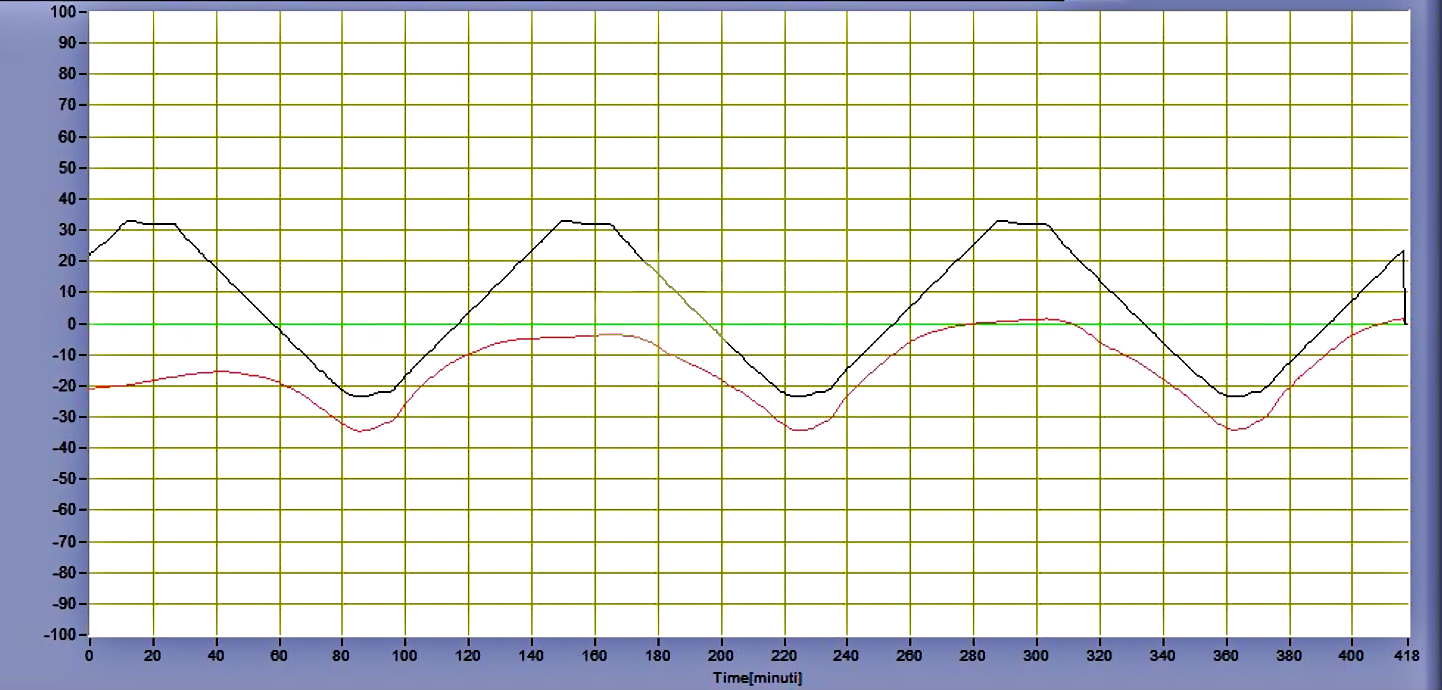}
\vspace{5pt}
\caption{BIL-type gas gap undergoing thermal cycling tests at INFN Frascati~(top). Recorded conditions within the test chamber~(bottom), illustrating temperature variations (black) between +30$^{\circ}$C and -20$^{\circ}$C across multiple thermal cycles, along with dew point variations (red), indicating humidity changes within the test chamber.}
\label{fig:frascati_thermal_cycling}
\vspace{15pt}
\end{figure}

A similar approach but a slightly different setup was used at the INFN Frascati to perform TCs on Phase II Barrel Inner Large (BIL) chambers. Figure~\ref{fig:frascati_thermal_cycling} shows BIL chambers being thermally cycled along with the recorded temperature and humidity variations during the process.

\section{Test Results} \label{sect:TestsResults}
Thermal cycling tests were conducted on five gas gaps at Cambridge and eight gas gaps at INFN Frascati, followed by a detailed assessment at the end of the test campaign. The BIS type gas gap samples, tested at Cambridge,  involved a total of 16 cycles, executed in the following order with each cycle followed by a leak measurement:
\begin{itemize}
    \item 1 cycle: from +25 to -20 $^{\circ}$C 
    \item 1 cycle: from -20 to +30 $^{\circ}$C
    \item 1 cycle: from -25 to +30 $^{\circ}$C
    \item 2 cycles: from -30 to +30 $^{\circ}$C
    \item 2 cycles: from -30 to +25 $^{\circ}$C
    \item 1 cycle: from -33 to +25 $^{\circ}$C
    \item 1 cycle: from -30 to +30 $^{\circ}$C
    \item 3 cycles: from -33 to +35 $^{\circ}$C
    \item 4 cycles: from -33 to +35 $^{\circ}$C
\end{itemize}

\begin{table}[H]
\vspace{0.8cm}
\centering
\caption{Comparison of pressure drop rates measured initially at CERN, re-evaluated at Cambridge before thermal cycling (TC), and after the TC campaign at Cambridge. The measurements were performed using a differential pressure manometer.}
\scalebox{0.95}{
\begin{tabular}{c|cc}
\hline
\textbf{Gas Gap} & \textbf{Before TC} & \textbf{After TC} \\
 & \textbf{(mbar/min)} & \textbf{(mbar/min)} \\
\hline
\hline
\textbf{BIS2A-01/23} &  No leak  & 0.016  \\
\textbf{BIS2A-08/23} &  0.015  & 0.018  \\
\textbf{BIS2A-11/23} &  No leak  & 0.006 \\ 
\textbf{BIS2A-12/23} &  No leak  & 0.013  \\
\textbf{BIS2A-28/23} &  0.013  & 0.020  \\
\hline
\end{tabular}}
\label{tab:leakRate}
\end{table}

The main finding of the QA tests carried out at Cambridge is that the TC protocol is found to have a notable ageing effect on the gas gaps, as shown in Table~\ref{tab:leakRate}. Since the measured leak rates in RPC gas gaps are directly related to pressure drop rates, the latter has been used as an alternative method for quantifying leak rates. As a result, these terms have been used interchangeably in the current work.

All gaps have developed a higher leak rate after the TC.
Most notably, gas gap BIS2A-28/23, one of the two gas gaps that were already exhibiting leaks at CERN, has developed quite a sizeable leak. 
This is a clear warning sign that this gap would not age well if deployed for the HL-LHC, and hence should never be installed.
On the other side, the gas gap BIS2A-08/23 that leaked initially, retained a relatively stable albeit notable leak rate. 
Finally, gas gap BIS2A-11/23, one of the three gaps that was not leaking initially, displayed some minor but below threshold leaks, while gaps BIS2A-01/23 and BIS2A-12/23 developed some leaks after TC. 
This requires additional investigation to pinpoint potential leakage sources. Initial measurements were taken over short intervals using a pressure manometer with a resolution of 0.01 mbar. 

\begin{figure}[H]
\centering
\includegraphics[width=8.5cm]{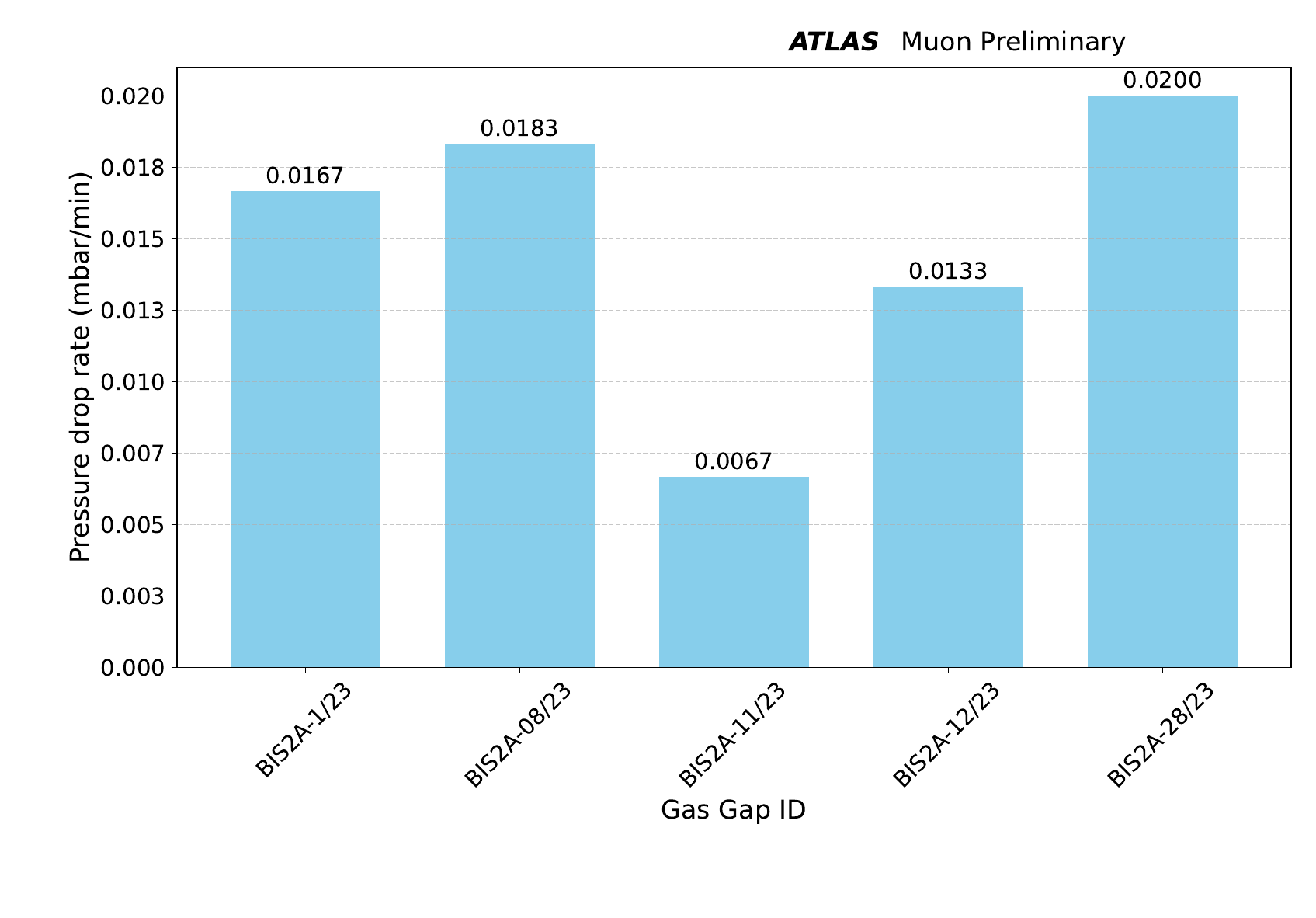}
\vspace{5pt}
\caption{Pressure drop rates for different gas gaps measured at Cambridge using a semi-automatic setup following thermal cycles.}
\label{fig:Camb_leakRates}
\vspace{15pt}
\end{figure}

For subsequent assessments, slightly longer-duration measurements were acquired with a semi-automatic setup, allowing for more precise monitoring. The semi-automatic setup used multiple cross-calibrated pressure and temperature sensors connected to a Raspberry Pi, enabling sensor control and data acquisition. The setup allowed better monitoring of gas gap conditions throughout the testing process. The results after the TC protocol are presented in Figure~\ref{fig:Camb_leakRates}. These measurements provide initial observations within the first few minutes after inflating the gas gaps to 5 mbar and are consistent with those obtained using the differential manometer (Table~\ref{tab:leakRate}). However, these should not be compared with the long-duration measurements presented later in this study. 
 
To put the TC tests into perspective, the Araldite glue, a two-component toughened methacrylate adhesive used in gas gap assembly, has a thermal expansion coefficient of 85–122 $\times$ 10$^{-6}$/$^{\circ}$C over the  -30 to +30 $^{\circ}$C range~\cite{Ref_Araldite_RSOnline, Ref_Araldite_hunt, Ref_Araldite}. By comparison, bakelite has a very low thermal expansion coefficient of 22 $\times$ 10$^{-6}$/$^{\circ}$C~\cite{Ref_ETB_Bakelite}, leading to stress at glue-bakelite interfaces and increasing rate of ageing when exposed to temperature variations.

\begin{figure}[H]
\centering
\includegraphics[width=8.5cm]{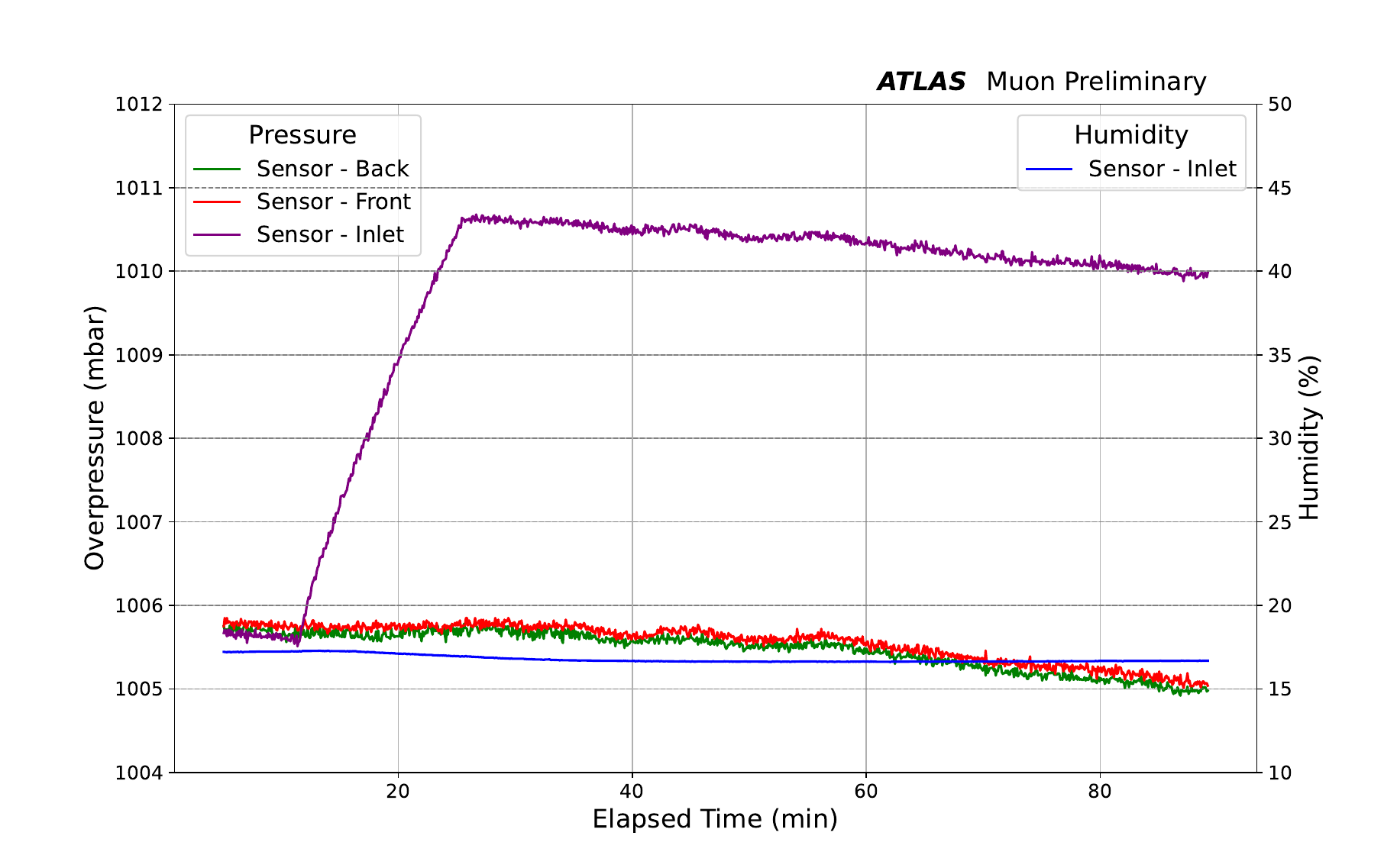}
\includegraphics[width=8.5cm]{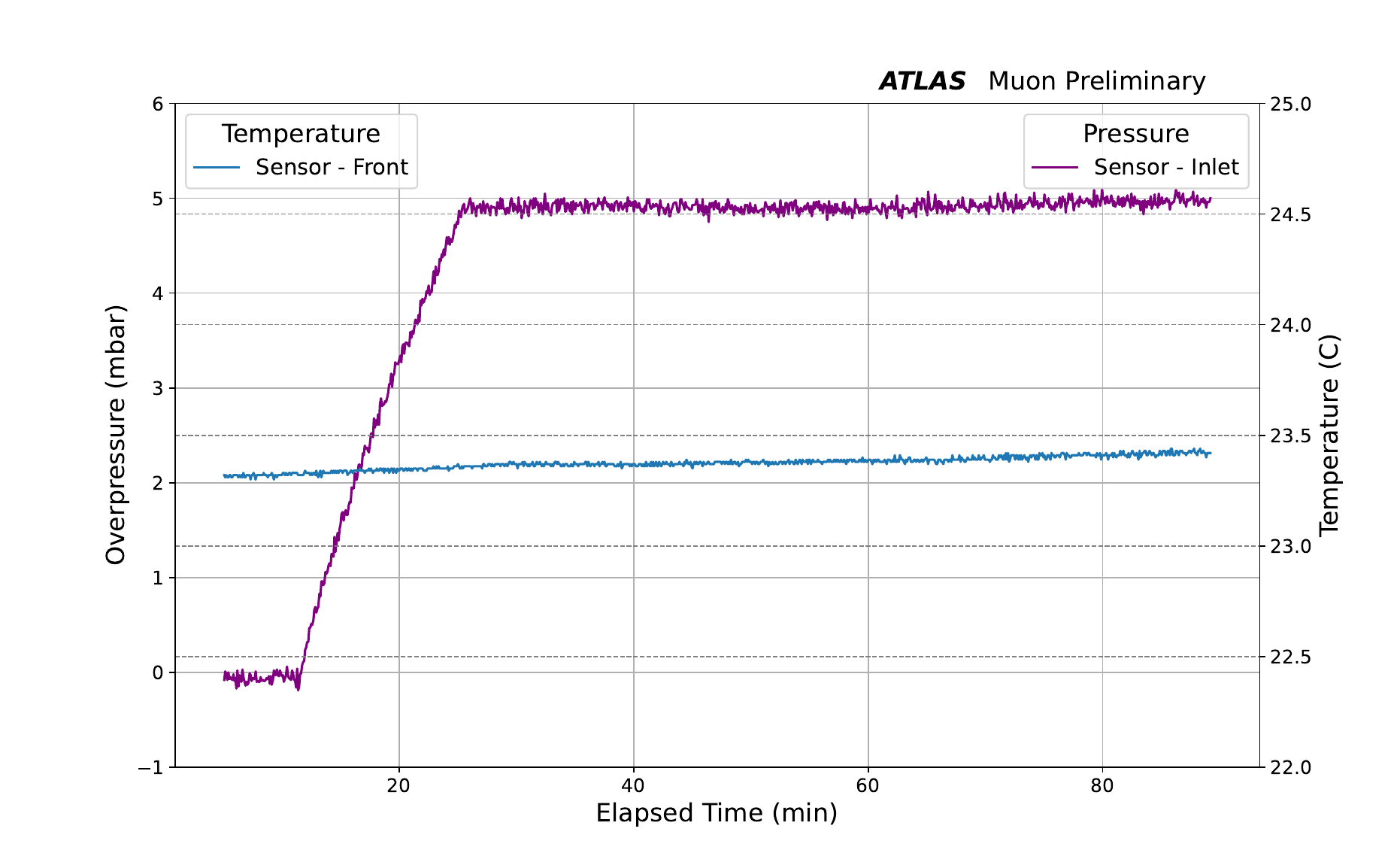}
\vspace{5pt}
\caption{Pressure drop over time for gas gap BIS2A-11/23 after 16 thermal cycles, measured using a semi-automatic setup with multiple sensors. (top) Absolute pressure values recorded over time, showing the initial inflated pressure of 5 mbar within the gas gap, ambient pressure from the Front and Back sensors, and internal pressure within the gas gap monitored by the Inlet sensor. Relative humidity is also shown for reference. (bottom) Pressure drop inside the gas gap over time, corrected for ambient pressure variations. The temperature trend is included, illustrating how a slight temperature increase affects the apparent pressure drop by masking it, a trend more clearly seen in Figure~\ref{fig:leaktests_SemiAutomatic_BIS2A_01_23}.}
\label{fig:leaktests_SemiAutomatic_BIS2A_11_23}
\vspace{15pt}
\end{figure}

As an example for a ``good'' gas gap, Figure~\ref{fig:leaktests_SemiAutomatic_BIS2A_11_23} presents the pressure drop for gap BIS2A-11/23 as a function of time after 16 TCs, with an initial overpressure of 5 mbar. The figure (top) shows concurrent ambient pressure fluctuations recorded alongside the pressure drop, while the figure (bottom) offers a corrected view of the pressure drop, adjusted for ambient pressure variations. Notably, the pressure remains largely stable, showing no signs of ageing, consistent with manual checks and measurements from CERN. A slight pressure decrease over time is observed; however, this is not immediately apparent in the figure due to a gradual temperature increase that obscures it. 

\begin{figure}[H]
\centering
\includegraphics[width=9cm]{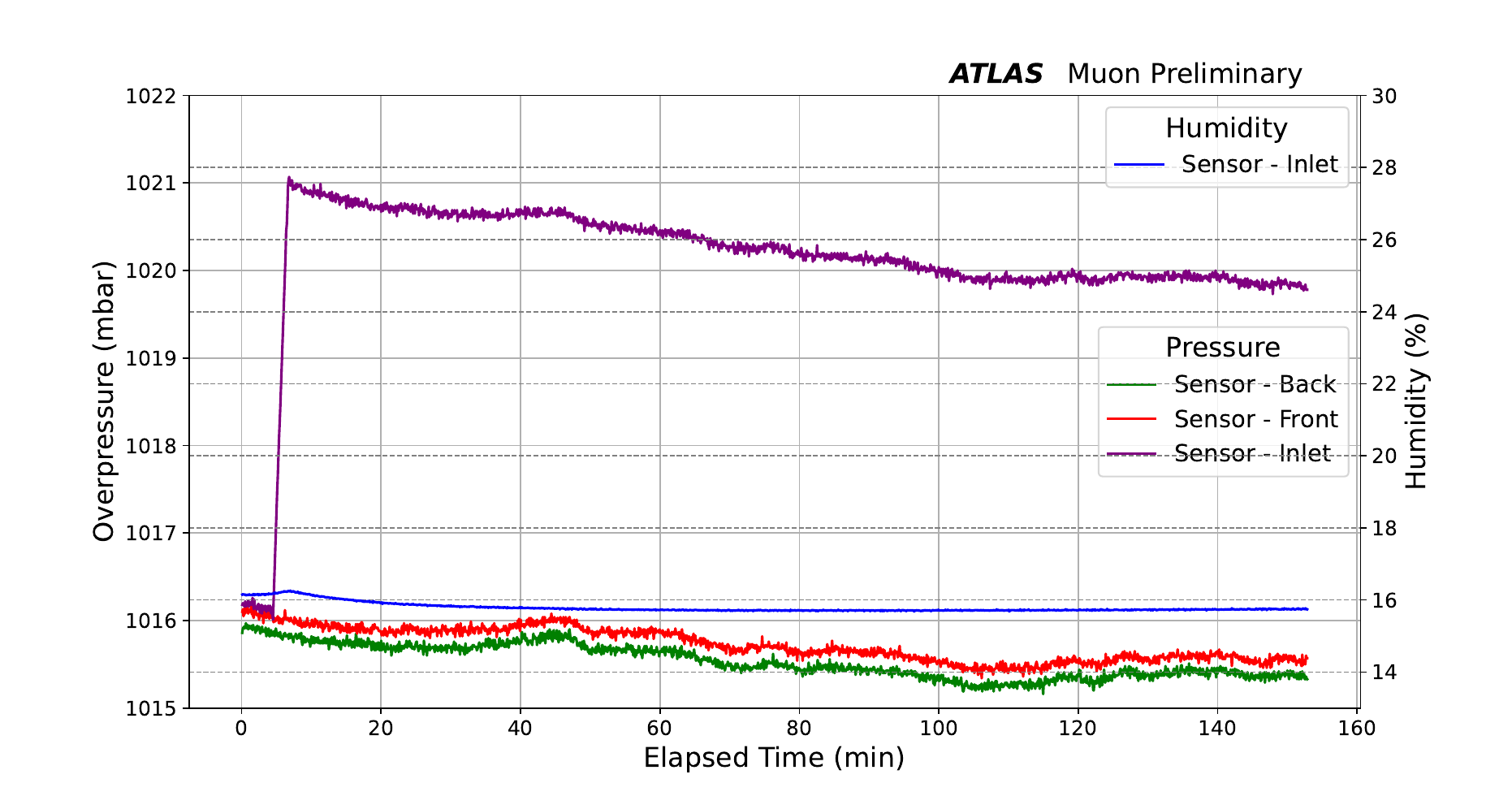}
\includegraphics[width=9cm]{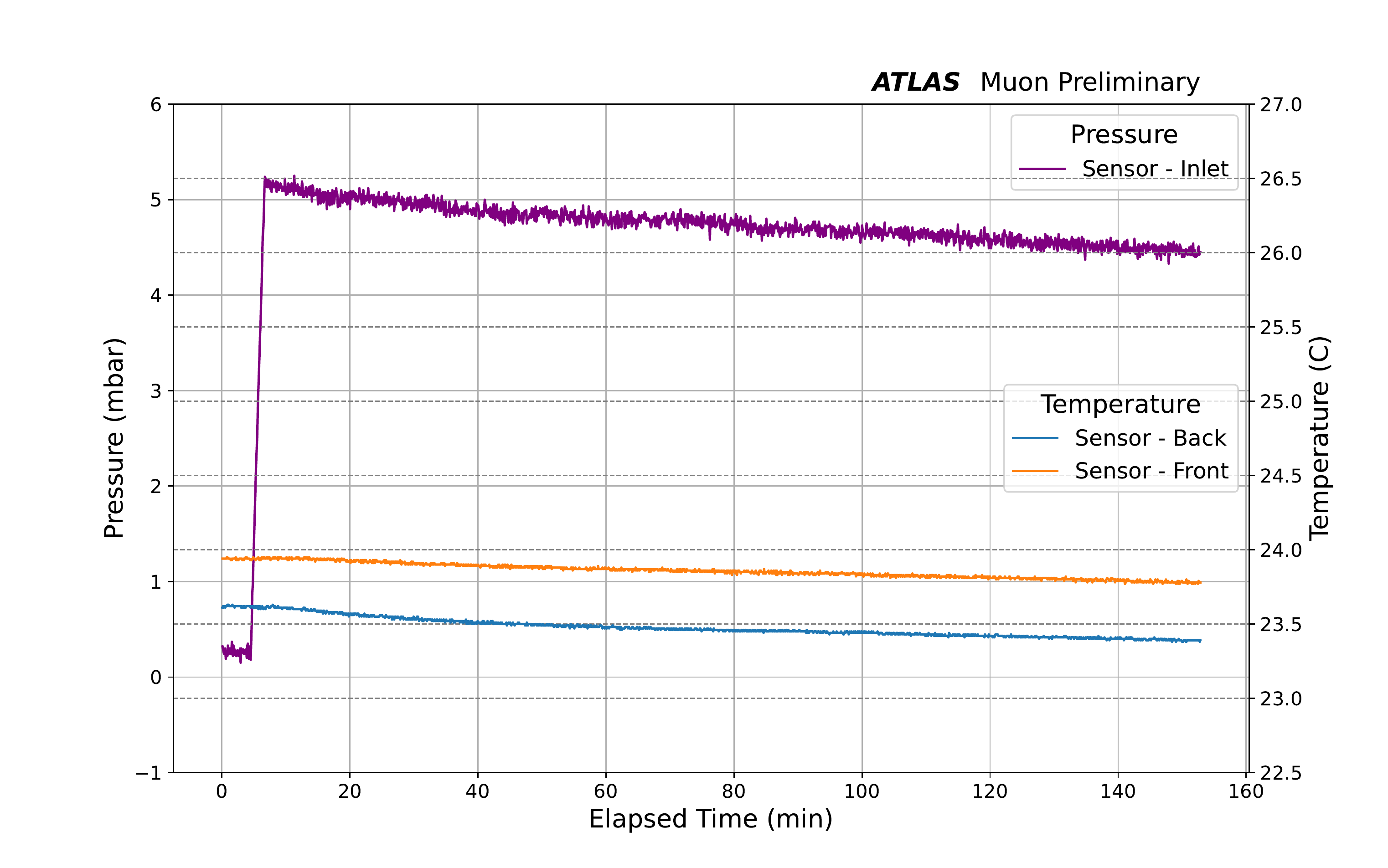}
\vspace{5pt}
\caption{Pressure drop over time for gas gap BIS2A-08/23 after 16 thermal cycles, recorded using the same setup and parameters described in Figure~\ref{fig:leaktests_SemiAutomatic_BIS2A_11_23}. This figure highlights the presence of leaks. In addition, relative humidity variations from the Inlet sensor are also included for reference in the figure at the top while the bottom figure includes temperature variations over a longer period recorded by Front and Back sensors, which track climate room conditions.}
\label{fig:leaktests_SemiAutomatic_BIS2A_08_23}
\vspace{15pt}
\end{figure}

To illustrate the behaviour of a ``bad'' gas gap, Figure~\ref{fig:leaktests_SemiAutomatic_BIS2A_08_23} shows pressure drop measurements for gap BIS2A-08/23 as a function of time after 16 TCs with stable temperature and humidity levels throughout. 
This gas gap exhibited a noticeable leak that worsened after thermal cycling.
It is interesting to note that the leakrate is fairly high in the first $\mathcal{O}(10)$ minutes, and stabilises somewhat afterwards. 
A potential explanation for this could be that the micropores that have opened are partially sealed by not fully polymerised linseed oil within the gap.

\begin{figure}[H]
\centering
\includegraphics[width=8.5cm]{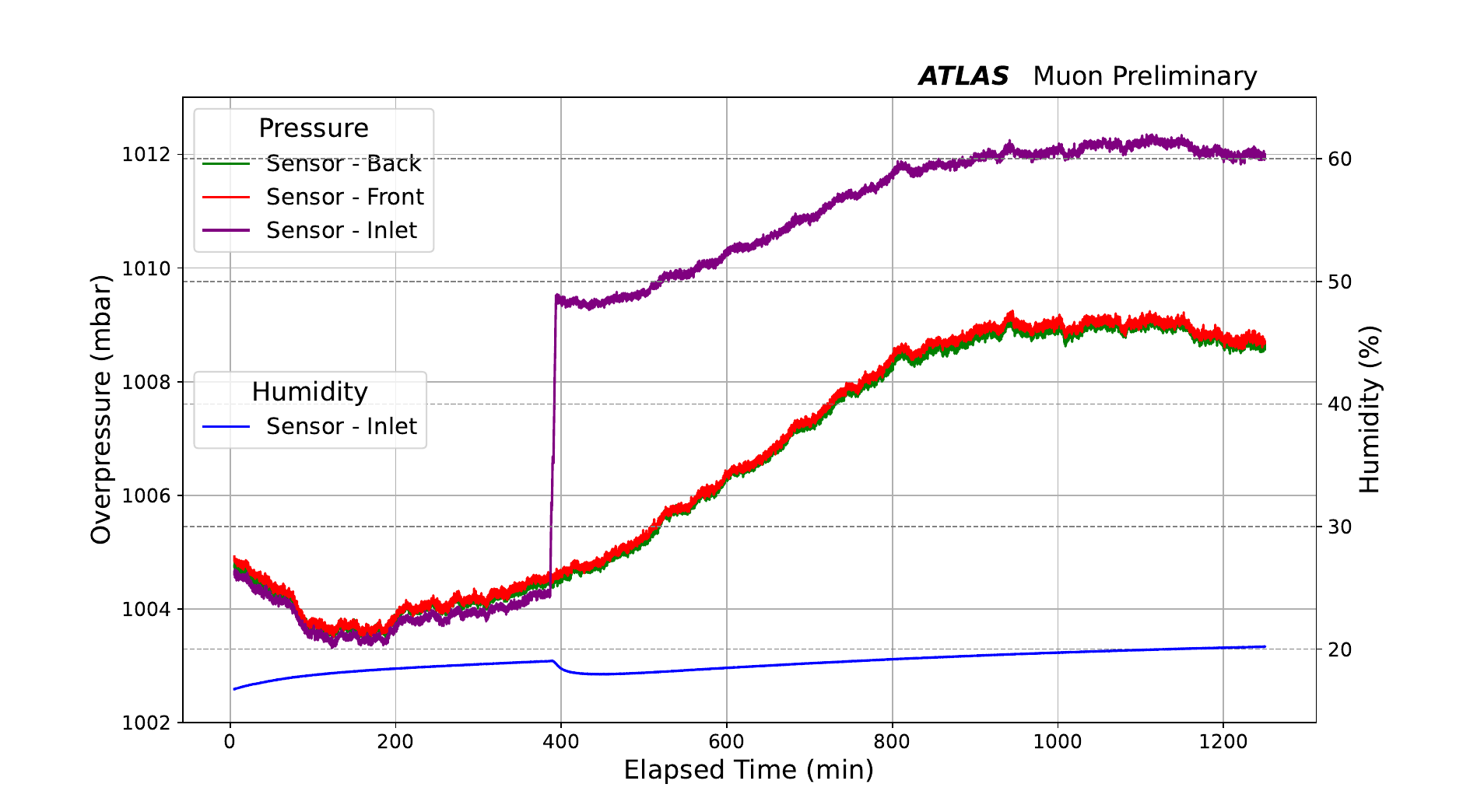}
\includegraphics[width=8.5cm]{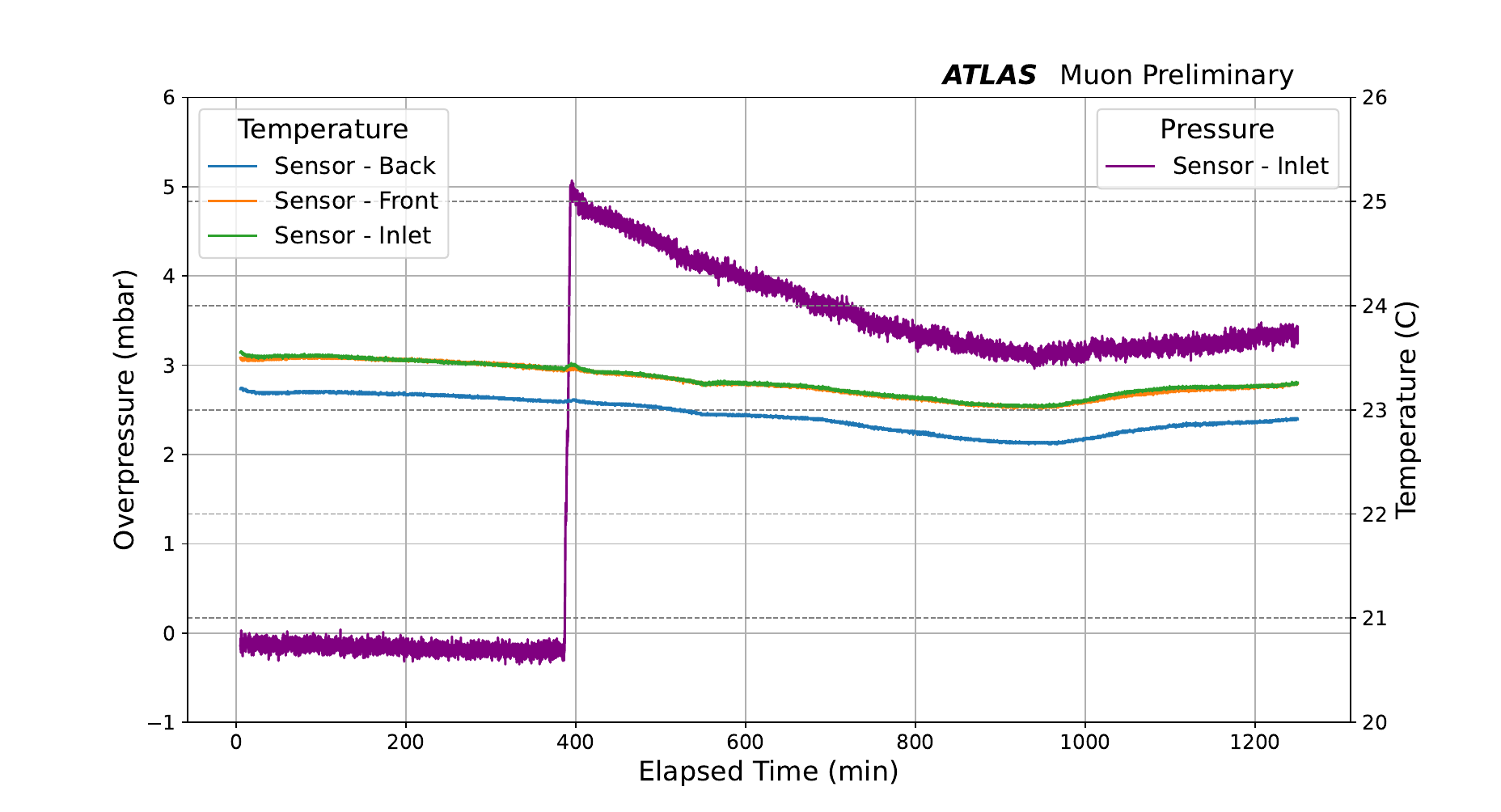}
\vspace{5pt}
\caption{Pressure drop over time for gas gap BIS2A-01/23 after 16 thermal cycles, recorded using the same setup and parameters described in Figure~\ref{fig:leaktests_SemiAutomatic_BIS2A_11_23}. This figure highlights threshold effects, suggesting that leaks may occur at higher overpressures (above 3 mbar), whereas the gas gap maintains adequate performance at lower overpressures (approximately 3 mbar or less).  Relative humidity variations are also included for reference in the figure at the top while the lower bottom one includes temperature variations over a longer period recorded by Front, Back, and Inlet sensors. The Front and Back sensors track external climate room conditions, while the Inlet sensor monitors internal temperature fluctuations within the gas gap. The slight increase in overpressure measurements toward the end is attributed to an overall temperature rise both inside and outside the gas gap.}
\label{fig:leaktests_SemiAutomatic_BIS2A_01_23}
\vspace{15pt}
\end{figure}

Another interesting result is presented in Figure~\ref{fig:leaktests_SemiAutomatic_BIS2A_01_23}, which depicts the leak rate for gas gap BIS2A-01/23 as a function of time after 16 TCs. It is apparent that over a period of more than 12h ambient pressure (top) and a subtle temperature differences (bottom) affect the leak rate. 
Specifically, a rapid initial decline in pressure is observed as the gas gap pressure drops from around 5 mbar, levelling off closer to 3 mbar. 
Similar to BIS2A-08/23 shown in Figure~\ref{fig:leaktests_SemiAutomatic_BIS2A_08_23}, the leak rate is significantly higher in the first $\mathcal{O}(10)$ minutes and decreases somewhat after that, although it remains intolerably high.
Additionally, a positive slope in the leak rate aligns with minor temperature increases in the bottom graph, suggesting some thermal sensitivity, although the trend stabilises briefly before rising again with temperature. 

The occurrence of a noticeable initial decline in the leak rate as the pressure decreases from 5 to 3 mbar for gap BIS2A-01/23 suggests that micropores or channels within the material may open or close only above certain pressure thresholds. As the pressure in the gas gap falls below a certain pressure (in the current case approximately 3 mbar), the leak rate stabilises, potentially indicating that these threshold-sensitive pathways are no longer active at lower pressures. These observed threshold effects, where certain pores open only at higher pressures (middle), are of particular interest. Identifying these threshold pressures is essential to optimise RPC performance in fluctuating pressures near operational levels. Such understanding could further enhance the RPCs' reliability, especially under varying pressure conditions close to their standard operating range.

The rapid pressure drop in the first $\mathcal{O}(10)$ minutes observed for gas gaps BIS2A-08/23 and BIS2A-01/23 may be attributed to a subtle effect: micropores that open up with the pressure increase are partially sealed again by not fully polymerised linseed oil or not fully cured glue within the gap. 
If confirmed, this would be a clear indication where the production process of the gas gaps should be improved.

An independent test campaign was undertaken at INFN Frascati to ensure comprehensive testing, including a systematic thermal cycling process with the following procedure

\begin{itemize}
    \item 5 cycles: from +10 to +30 $^{\circ}$C
    \item 4 cycles: from +0 to  +30 $^{\circ}$C
    \item 3 cycles: from -10 to +30 $^{\circ}$C
    \item 3 cycles: from -20 to +30 $^{\circ}$C
\end{itemize}

After each set of thermal cycles was completed, leak rate assessments were conducted to monitor the integrity of the gas gaps, starting from an overpressure of 3~mbar. Remarkably, these assessments revealed no leaks, indicating that the gas gaps maintained their gas tightness throughout the testing process. However, minor leaks were occasionally detected at Frascati in the gas gap inlets and outlets, even without thermal cycling, emphasizing the importance of further testing and supporting the inclusion of thermal cycling as an additional quality control measure.

The differences between the Cambridge and Frascati results may be attributed to subtle but important differences in the testing protocol:
\begin{itemize}
\item 
First, the leak measurements were performed at an overpressure of 5 mbar at Cambridge versus 3 mbar at Frascati. As the results from of gap BIS2A-01/23 shown in Fig.~\ref{fig:leaktests_SemiAutomatic_BIS2A_01_23} indicate, micropores or other effects that become measurable above 3 mbar are likely to play an important role in this comparison.
\item 
Furthermore, thermal cycles spanned a maximum temperature range of $-20^\circ$ and $+30^{\circ}$C at Frascati, with only three cycles performed in this maximum temperature excursion. By contrast, the tests at Cambridge involved a broader temperature range, from $-33^\circ$ to $+35^{\circ}$~C; seven cycles were completed, including additional cycles at conditions closer to the maximum temperature excursions. 
\end{itemize}
%
%

\section{Fluorescent Tracer Tests} \label{sect:AdditionalTests}
\begin{figure}[H]
\centering
\includegraphics[width=9.3cm, height=7.3cm]{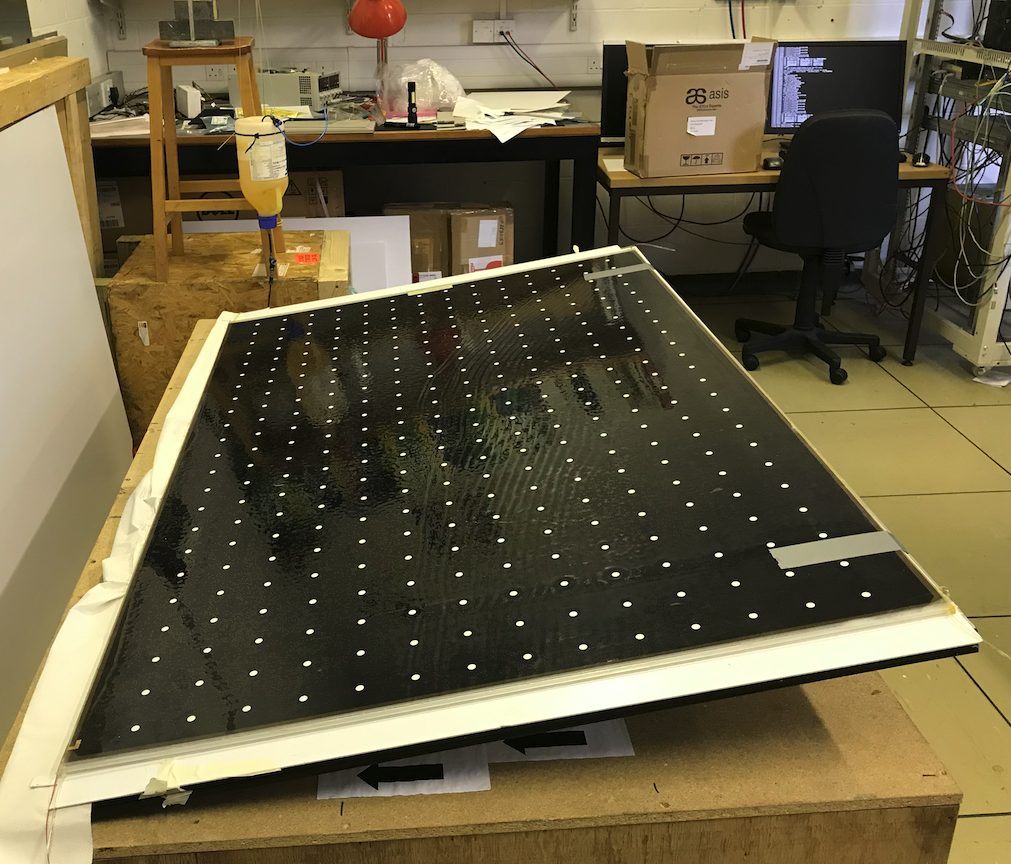}
\includegraphics[width=5.1cm, height=4.5cm]{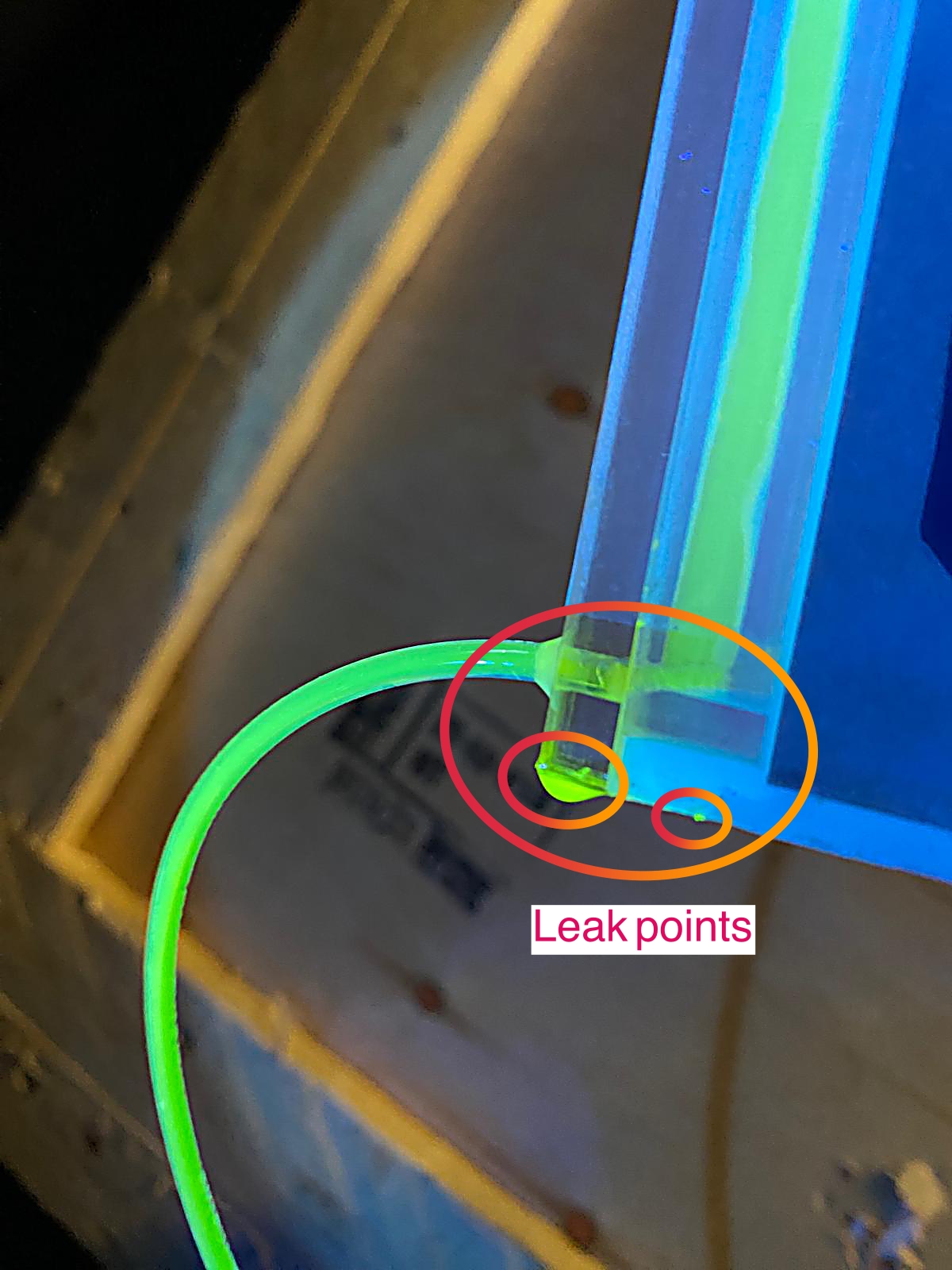}
\includegraphics[width=5.1cm, height=4.5cm]{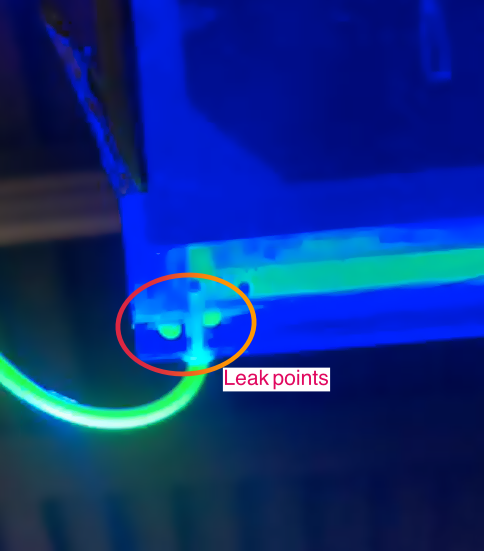}
\vspace{5pt}
\caption{(top) A gas gap being filled with the fluorescent tracer solution using a bottle, with the height of the bottle gradually raised from the bottom. (bottom) Gas gaps filled with the fluorescent tracer solution illuminated with UV light (external lights switched OFF), illustrating the process of leak detection. The concentration and path of the tracer (in green) highlighted in these images indicate the presence of leaks or potential defects near the gas inlet/outlets and within the gas distribution manifold.}
\label{fig:F_testPics_cambride_frascati}
\vspace{15pt}
\end{figure}

After initial assessments revealed leaks in the gas gaps in one of the test campaigns, a gas sniffer -- Heated Diode Refrigerant Leak Detector (Model DR58)~\cite{Ref_Gas_Sinf_DR58} -- was employed to roughly locate these leaks. To identify the critical points where the leaks occurred more accurately, additional tests were conducted using a solution with a fluorescent tracer `Fluorescein'. This solution was prepared by mixing Fluorescein powder with isopropyl alcohol, creating a detectable medium that could highlight effectively the path of leaks when illuminated with ultraviolet (UV) light. The two gas gaps that were considered for this test at Cambridge were BIS2A-28/23 and BIS2A-08/23, as these had higher leak rates, as depicted in Figure~\ref{fig:Camb_leakRates}. The gas gaps were filled with the fluorescent tracer solution, and a UV torch was used to illuminate the tracer within the gas gaps allowing for the visualisation of the tracer's path, as shown in Figure~\ref{fig:F_testPics_cambride_frascati}. Under UV light, any critical areas such as micro-cracks, weak joints, or potential leak sites would be illuminated, revealing regions where the fluorescent light was more concentrated or where it visibly seeped out of the gas gap. Similar tests were also conducted at Frascati. The results from the tracer tests were consistent with those from the sniffer leak detection tests and assessments, identifying specific leaks and weak points that were primarily associated with the gas distribution manifolds within the gas gaps. These issues were promptly followed and addressed by the ATLAS Muon Community.


\section{Summary and Outlook} \label{sect:Summary}
The Phase II upgrade of the ATLAS Muon Spectrometer will introduce several hundred next-generation RPCs. For the first time, thermal cycling was employed as a quality assurance technique to evaluate the gas tightness of RPC gas gaps. This new approach, inspired by the ATLAS ITk module quality control, specifically aimed to simulate potential long-term mechanical stress and ageing by focusing on critical interfaces such as those between the bakelite-adhesive layers.

The current work, therefore, marks another step in advancing the quality assurance and quality control process for RPC gas gaps, providing promising evidence of the effectiveness of thermal cycling in revealing structural weaknesses. Some gas gaps that initially tested leak-tight developed leaks after thermal cycling, while pre-existing leaks in others worsened. These observations highlight the role of thermal cycling in identifying structural vulnerabilities, particularly under temperature variations. The study also revealed that leaks originated near specific regions, prompting targeted design improvements to mitigate these weaknesses.


An additional important finding was the identification of pressure-dependent leak behaviour, where gas gaps showed transient leaks at comparatively higher overpressure but stabilized at lower pressures. This phenomenon, likely caused by micropores or material channels opening under specific conditions, introduces an important aspect for further investigation. Understanding these threshold effects could enhance the reliability of RPCs under fluctuating operational pressures and ensure stable operation on time scales of the HL-LHC.

Approximately 25\% of the BIS and BIL gas gaps have been tested so far, with a fraction subjected to thermal cycling. Notably, leaks were predominantly localized around the gas inlets and outlets, highlighting these regions as critical weak points. These findings helped to reassure and support the ATLAS team’s pre-existing plans for the redesign of the gas inlets/outlets and distribution manifolds to address the identified vulnerabilities. Gas gaps incorporating the revised design are now in production, and future studies might focus on validating these improvements. Additionally, the adhesive materials used in gas gap assembly are undergoing further scrutiny to assess their long-term durability. With ongoing refinements in design and materials, these results contribute to ensuring the long-term reliability of the RPC system under HL-LHC conditions.


\section*{Acknowledgements}
We extend our gratitude to the Department of Applied Mathematics and Theoretical Physics (DAMTP) at the University of Cambridge for granting us access to their climate chamber facility. We are especially grateful to Prof. Stuart Dalziel and Dr. Mark Hallworth for their support and assistance in facilitating part of this work at DAMTP. We also thank Dr.~Bart Hommels for his valuable insights on thermal QA/QC. 
We extend our sincere gratitude to our technical team, Richard Shaw and Gaurav Kumar, for their dedicated help and assistance throughout this work, and to the technical teams at DAMPT and Frascati.
\section*{Copyright}
Copyright 2025 CERN for the benefit of the ATLAS Collaboration. Reproduction of this article or parts of it is allowed as specified in the CC-BY-4.0 license.

\end{document}